\documentclass[12pt]{article}
\usepackage[authoryear, round]{natbib}
\usepackage{graphicx}
\usepackage{amsmath}
\def\be{\begin{equation}}
\def\ee{\end{equation}}
\usepackage{float}
\usepackage{setspace}

\title{Adaptive Model-Switch Thermodynamic Integration: A Bayes Factor Estimator}
\author{Amanda Malnati, Megan Kralj, and Ken D. Olum\\
Institute of Cosmology\\Department of Physics and Astronomy\\Tufts University, Medford, MA 02155, USA}

\date{}

\providecommand{\keywords}[1]
{
  \small	
  \textbf{\textit{Keywords---}} #1
}

\begin{document}

\maketitle

\begin{abstract}
 Bayes factors are a widely used model comparison criterion, but are often quite challenging and computationally expensive to estimate well. We propose a variation of the model-switch thermodynamic integration method for estimating Bayes factors that decreases the variance of the estimator and essentially eliminates integration error.  We choose samples whose density in $\beta$ is proportional to the standard deviation of previously drawn samples.  Rather than averaging many samples at each $\beta$, we use a very dense set of $\beta$ values, collecting one sample at each in order of increasing $\beta$ and one sample at each in order of decreasing $\beta$ to minimize ``thermic lag''.  We study our procedure with a toy model (based loosely on some signal recovery problems).  In this model we find that adaptive sampling gives us a 2.7-fold decrease in the spread of the estimated Bayes factor.  We test our procedure both drawing directly from the posterior distribution and using a Markov Chain Monte Carlo sampler. 
\end{abstract}


\section{Introduction}

Across fields, determining which hypothesis best fits the data is a common problem known as model selection. Suppose we have collected some observational data $d$. We wish to determine which of several candidate models, $M_A$, $M_B$, ..., each with its own set of tunable parameters, $\theta = \{ \theta_1, \theta_2, \theta_3, ...\}$, best explains the data without being overly complex. A wide range of comparison criteria have been developed (see e.g. \citet{kadane_lazar_2004}), but in this paper, we focus on the Bayesian approach to this problem.

In the Bayesian framework, a model is defined by a likelihood function $L(d|\theta)$, which describes how probable the observed data are given the model parameters $\theta$, together with a prior distribution $\pi(\theta)$ that encodes our assumptions and uncertainties about those parameters before seeing the data.

We combine these using Bayes's Theorem to form the unnormalized posterior distribution of a model,
\be
q(\theta) = L(d|\theta) \pi(\theta)\,,
\ee
which we then normalize,
\be
\label{eqn: model posterior}
P(\theta|d) =  \frac{q(\theta)}{Z}\,,
\ee
where $Z$ is the normalizing constant, analogous to the partition
function of statistical mechanics,
\be
Z = \int q(\theta) d\theta \,.
\ee
The distribution $P(\theta|d)$ is the normalized probability of parameter values given the data. This form of Bayes's Theorem allows us to estimate which parameter values are most consistent with the data. 

However, parameter estimation alone does not address the question of
model selection. To compare models, we instead consider the marginal
likelihood, or evidence, of a model, which is the just normalizing
constant $Z$, which averages the likelihood $L(d|\theta)$ over all
parameter values weighted by the prior $\pi(\theta)$.  The marginal
likelihood gives an average fit of the model rather than evaluating
the fit at one specific point in parameter space.

Within the Bayesian framework, a widely used model comparison criterion is the Bayes factor (BF) \citep{jeffreys_1935, jeffreys_1998}. The BF is a ratio of the marginal likelihoods of two competing models A and B, 
\be
BF_{B,A} = \frac{Z_B(d)}{Z_A(d)} = \frac{\int L_B (d|\theta) \pi_B(\theta) d\theta}{\int L_A (d|\theta) \pi_A(\theta) d\theta}\,,
\ee
where a BF of 1 indicates equal evidence for both models, and values greater than 1 show preference for model B.
As a ratio of the probabilities of the observed data for all possible parameter values, the BF naturally balances both the accuracy and the complexity of the models \citep{jeffreys_1935, jeffreys_1998}.

In practice, computing marginal likelihoods is challenging, as it requires integrating over the full parameter space of each model. In all but the simplest cases, it cannot be done analytically. For a review of techniques, see \cite{friel_wyse_review}.  One useful approach is thermodynamic integration (TI) or path sampling \citep{friel_pettitt_2008,lartillot_philippe_2006}, which estimates the marginal likelihood by introducing a continuous path of intermediate distributions indexed by a parameter $\beta \in [0,1]$

The method is motivated by an analogy to statistical mechanics, where one computes the difference in Helmholtz free energy between two states by integrating along a path connecting them \citep{kirkwood_1935}. In the context of marginal likelihood estimation, the path to be integrated over connects the prior and posterior of a single model. To further simplify the problem \citet{friel_pettitt_2008} suggest combining the two integrals needed to estimate the BF into one model-switch formulation, generating a path that connects the posterior distributions of two competing models directly, allowing the BF to be estimated from a single integration.

The TI method requires us to draw samples from the distributions along the TI path. Markov chain Monte Carlo (MCMC) methods allow us to explore parameter space and do Bayesian inference with high dimensional models while avoiding an integral over all of parameter space \citep{Geyer1992}.  We can then approximate the integral by sampling from intermediate distributions at a discrete set of $\beta$ values, which we will refer to as a $\beta$ schedule.  This is related to the idea of a temperature ladder, with $\beta = 1/T$.

The practical performance of TI strongly depends on the choice of the $\beta$ schedule. If the schedule is too coarse in regions where the integrand changes rapidly, the discretization error of the integral approximation can become large. If too much effort is spent on regions that are easy to estimate, computational resources are used inefficiently. 

The rest of this article is organized as follows. In Section~\ref{sec: Thermodynamic Integration} we describe the single and model-switch TI methods, and discuss the errors associated with these methods. In Section~\ref{sec: The beta Schedule} we propose an adaptive model-switch TI scheme that iteratively distributes sampling locations according to the observed variability of the log-posterior ratio along the path. Then in Section~\ref{sec: Toy Model} we describe a toy model with known BF, and in Section~\ref{sec: Adaptive Model-Switch TI with the Toy Model} we test the adaptive model-switch method with this toy model and compare it against non-adaptive schedules, with particular attention to variance reduction and the concentration of samples near difficult regions of the path. Section~\ref{sec: Conclusion} summarizes our findings.

\section{Thermodynamic Integration} \label{sec: Thermodynamic Integration}
\subsection{Basic Formalism}

In thermodynamic integration, we integrate over a path in the space of unnormalized posterior functions from some $q_0(\theta)$, which
may be just the prior, to some $q_1(\theta)$.  The path is given by the ``power posterior'' \citep{friel_pettitt_2008},
\be
q_\beta(\theta) = q_0(\theta)^{1-\beta} q_1(\theta)^{\beta}\,.
\ee
Each $q_\beta$ has a normalizing constant
\be
Z_\beta = \int q_\beta(\theta) d\theta \,.
\ee
The logarithmic derivative of $Z_\beta$ is
\be
\label{eqn: derivative TI}
\frac{d}{d\beta} \ln Z_\beta =
\frac{1}{Z_\beta}\int \frac{dq_\beta(\theta)}{d\beta} d\theta
= \frac{1}{Z_\beta}\int \frac{d\ln q_\beta(\theta)}{d\beta}
q_\beta(\theta)  d\theta
= E_\beta \left[ \frac{d}{d\beta}\ln q_\beta(\theta) \right]
\ee
Where $E_\beta[...]$ denotes the expectation value with respect to
$P_\beta(\theta)=q_\beta(\theta)/Z_\beta$. By integration we can now find the Bayes factor,
\be
\label{eqn: fund. TI}
\ln \frac{Z_1}{Z_0} = \int_0^1 E_\beta \left[ \frac{d}{d\beta}\ln q_\beta(\theta) \right] d\beta\,.
\ee
This is the fundamental equation of TI.  For more details see \citet{lartillot_philippe_2006,aponte_2021,friel_pettitt_2008}.

\subsection{Single Model} 

Single model TI has been shown to be a reliable method for estimating
the marginal likelihood in many fields including biology and medicine
\citep{lartillot_philippe_2006,friel_pettitt_2008,xie_et_al_2010}). For
single model, the two ends of the path are the prior and the posterior, so we
let $q_0(\theta) = \pi(\theta)$ and $q_1(\theta) = L(d|\theta) \pi(\theta)$.
Plugging directly into the fundamental TI equation (\ref{eqn:
  fund. TI}), and noting that $Z_0 = \int \pi(\theta) d\theta = 1$, we
are able to simplify the equation to
\be
\ln Z_1 = \int_0^1 E_\beta \left[ \ln L(d|\theta) \right] d\beta
\ee
where $Z_1 = \int L(d|\theta) \pi(\theta) d\theta$ is the marginal likelihood of our one model.

At each value of $\beta$, the expectation value can be approximated by MCMC sampling from the power posterior distribution, giving
\be
\label{eq:MCMCTI}
   \widehat{Z_1} = \exp\int_0^1 \frac{1}{n} \sum_{i=1}^n \ln L(d|\theta_i)  d\beta \,,
\ee
where the $\theta_i$ are $n$ independent samples drawn from the distribution $P_\beta(\theta)$.  MCMC makes it possible to draw the samples knowing only the unnormalized distribution $q_\beta(\theta)$.
The integral over $\beta$ can then be numerically evaluated with standard techniques, such as the trapezoid rule, using samples from $q_\beta(\theta)$ at discrete values of $\beta$ determined by the $\beta$ schedule.

\subsection{Model-Switch Thermodynamic Integration}
\label{subsec: MSTI}
In the model-switch formulation \citep{lartillot_philippe_2006},
instead of taking a path from the prior to the posterior of each
model, we use a single path between the unnormalized posteriors of
Model A and Model B to directly estimate the BF.  Thus we let
\be
   \label{eq:ms_unnormalized} q_0(\theta) = L_A(d|\theta)
   \pi_A(\theta) \qquad \text{and} \qquad q_1(\theta) = L_B(d|\theta) \pi_B(\theta)\,.
\ee
The parameter vector $\theta$ contains all parameters used by either model.  If a parameter is not used by one of the models, it does not enter into that model's likelihood and we will generally use the same prior as the model in which the parameter is used.\footnote{An improvement can be made by the use of pseudo-priors \citep{vitoratou_ntzoufras_2016} or with the techniques of \citet{oates_papamarkou_girolami_2016}} 
For the A-to-B path used here, we can again plug into the fundamental TI equation (\ref{eqn: fund. TI}) to get
\be
\ln \frac{Z_1}{Z_0} = \int_0^1 E_\beta \left[ \ln \frac{L_B(d|\theta) \pi_B(\theta)}{L_A(d|\theta) \pi_A(\theta)} \right] d\beta\,.
\ee
Thus we have directly computed the BF with a single thermodynamic integration,
\be
    \label{eq:ms_ti}
    \mathrm{BF}_{B,A}^{\mathrm{TI}} = \exp \int_0^1  E_{\beta}\left[\ln\frac{L_B(d|\theta) \pi_B(\theta)}{L_A(d|\theta) \pi_A(\theta)}\right]  d\beta\,.    
\ee
The expectation value can be approximated by an average over $n$ samples,
\be
    \label{eq:ms_ti_estimate}
    \widehat{\mathrm{BF}}_{B,A}^{\mathrm{TI}} = \exp \int_0^1 \frac{1}{n} \sum_{i=1}^n\ln\frac{L_B(d|\theta_i) \pi_B(\theta_i)}{L_A(d|\theta_i) \pi_A(\theta_i)} d\beta \,,
\ee
where the samples are drawn using MCMC from $q_\beta(\theta)$ at a given value of 
$\beta$.
We can then approximate the integral over $\beta$ using samples taken at a discrete set of 
$\beta$ values as in the single model scenario.  Since model-switch TI requires only a single integration, it involves fewer sources of error than estimating the two marginal likelihoods separately, especially if the models are very similar.
\subsection{Errors in Thermodynamic Integration}

TI is affected by both numerical integration and sampling errors. Numerical integration (discretization) error arises because the $\beta$ integral is evaluated on a finite schedule.  If our spacing is too coarse, we will miss sharp features in the integrand. Sampling error arises because the expectation at each $\beta$ is estimated from a finite sample rather than being evaluated exactly. In the limit $\Delta \beta \to 0$, the discretization error vanishes, while the sampling error can be reduced by increasing the number of samples.

In practice, without knowing the analytical solution, it is difficult to determine the best number of intermediate distributions and samples under finite computational cost. In the MCMC context, the sampling error is further compounded by autocorrelation, which reduces the effective number of independent draws, and by possible bias if the sampler moves through $\beta$ too quickly to equilibrate to the local target distribution. A practical schedule must therefore balance path resolution, sampling effort, and local mixing. 

\section{The $\beta$ Schedule}
\label{sec: The beta Schedule}

\subsection{Predetermined $\beta$ Schedule}
The simplest $\beta$ schedule would be a discrete linear spacing, with samples at each distribution drawn independently and in parallel. \citet{lartillot_philippe_2006} build upon this with a ``continuous'' linear spacing that they call the quasistatic method, in which a single sampler takes steps of $\Delta\beta = 0.01$, and draws $10^3$ samples at each step. \citet{xie_et_al_2010} suggest a skewed schedule for which the $\beta$ values are evenly spaced quantiles from a Beta($\alpha$, 1) distribution, finding the best results for $0.2 < \alpha < 0.4$. This generalized the \citet{friel_pettitt_2008} method which is specific to $\alpha = 0.25$. These methods place more samples near $\beta = 0$ in single model TI, where the integrand is often quite steep. Specifically in the context of the model-switch framework, \citet{lepage_2007} propose spacing $\beta$ with a sigmoidal function, so that more $\beta$ are placed towards $\beta=0$ and $\beta=1$, where the integrand is evolving more rapidly and is prone to larger discretization error. 

\subsection{Adaptive $\beta$ Schedule}
Predetermined schedules allocate samples according to a fixed rule specified before the calculation begins. Although convenient, such schedules can be inefficient in regions where the integrand varies sharply with $\beta$, because they do not incorporate information from the realized integrand itself.  \cite{friel_hurn_wyse_2013} propose an adaptive method whose goal is to improve integration error, but we will argue in the next subsection that integration error can be essentially eliminated, so we concentrate here on sampling error.

In practice, model-switch integrands can have some regions of $\beta$ that are smooth and easy to estimate, while others can exhibit high variability or rapid change, thus dominating uncertainty in the final Bayes-factor estimate. This motivates an adaptive strategy in which the sampling density is updated using information gathered from previous runs.

The scheme we outline in the following section is an iterative method for determining how to distribute samples across $\beta$, designed to build on an initial chosen $\beta$ schedule. It originates with the concept of \citet{neyman_1934} allocation.\footnote{See \citet{lee_jahanmahin_pal_york_2025} for a similar application of Neyman allocation in the context of alchemical free energy calculations.  An analysis of optimal sampling strategy was done by \citet{GelmanMeng} (see also \citet{CalderheadGirolami}), but they drew samples from a distribution $p(\beta)$ rather than using fixed set of $\beta$.  The variation of the set of $\beta$ leads to a higher variance and a somewhat different optional sampling density.} The concept is simple: to minimize the variance in our estimator, the density of samples taken in a local region of $\beta$ should be proportional to the local standard deviation of the integrand, as we now show.

Suppose we have a set of positions $\beta_j$, and we draw $n_j$ samples at $\beta_j$.  Our estimate of the log BF is then
\be
    \label{eq:ms_ti2}
    \widehat{\ln\mathrm{BF}}_{B,A}^{\mathrm{TI}} = \sum_j w_j \frac{1}{n_j} \sum_i\ln\frac{L_B(d|\theta_i) \pi_B(\theta_i)}{L_A(d|\theta_i) \pi_A(\theta_i)} \,,
\ee
where the $w_j$ are the weights of our integration formula.  For the
trapezoid rule, $w_j$ is just the width in $\beta$ represented by the
samples at $\beta_j$.  Let $\sigma_j^2$ be the variance of the
distribution at $\beta_j$.  (This is also the slope of
the mean at $\beta_j$.  See Appendix~\ref{sec: FDT}) The expected variance of the sample mean in group $j$ is then $\sigma_j^2/n_j$, so the total variance of (\ref{eq:ms_ti2}) is
\be\label{eq:var1}
\sigma^2_{\mathrm{total}} =
\sum_j\frac{w_j^2\sigma_j^2}{n_j}\,.
\ee
Suppose our computational resources limit us to some total number of samples $N =\sum n_j$.
To minimize the variance in our BF estimate subject to this constraint, we vary the $n_j$ using a Lagrange multiplier, and require
\be
0 = 
\frac{d}{dn_j}
\left[
\sum_{j=1}^M \frac{w_j^2\sigma_j^2}{n_j}
+ \lambda \left( \sum_{j=1}^M n_j - N \right)
\right]
= -\frac{w_j^2\sigma_j^2}{n_j^2} + \lambda \,,
\ee
so $n_j \propto w_j\sigma_j$: the number of samples should be proportional to the weighted standard deviation.  If the weight is given by the range of $\beta$ represented, the density of samples is proportional to the standard deviation,
\be
n_j/w_j \propto \sigma_j \,.
\ee

\subsection{One sample per $\beta$}
It appears that there is a tension between using many samples at each $\beta_j$ to reduce the sampling error and using many different $\beta_j$ to reduce the integration error.  But as we now show, this is not the case.  It is always better to spread out the samples and get one sample at each $\beta$.

From (\ref{eq:var1}), the contribution of the $n_j$ samples at $\beta_j$ to the variance of our BF estimate is
\be\label{eq:varstacked}
w_j^2\sigma_j^2/n_j \,.
\ee
Now suppose instead we spread out these samples into $n_j$ different $\beta$ values, which we will denote $\beta_{jk}$ with $k=1\ldots n_j$.  At each $\beta_{jk}$ we take one sample with weight $w_j/n_j$.  The variance due to these $n_j$ samples is now
\be\label{eq:varspread}
     \sum_{k=1}^{n_j} \frac{w_j^2}{n_j^2} \sigma_{jk}^2 \,.
\ee
If the range of $\beta$ over which we spread the samples is small, the sum of $\sigma^2_{jk}$ is approximately $n_j\sigma^2_j$, so~(\ref{eq:varspread}) is the same as~(\ref{eq:varstacked}). Thus spreading the samples across nearby $\beta$ values approximately preserves the sampling variance while producing a finer quadrature grid and reducing discretization error.

\subsection{General Procedure}
\label{sec: General Procedure}
We thus recommend the following procedure:
\begin{enumerate}
    \item Draw $N$ samples from the power posterior
    $P_\beta(\theta)$, with an initial $\beta$ schedule of size $N$. The simplest choice would
    be a linear schedule.  Give each sample a weight $w_i$, e.g., by the trapezoid rule.

    \item Compute the log posterior ratio of the two models,
    \be
    \ln \left[\frac{L_B(d|\theta_i)\pi_B(\theta_i)}{L_A(d|\theta_i)\pi_A(\theta_i)}\right]\,,
    \ee
    for each sample $i$.

    \item Divide the samples into successive groups in $\beta$ containing equal numbers of samples and find the standard deviation of the log posterior ratio in each group.

    \item For a new set of samples $N_{\mathrm{new}} (> N)$, the number of samples allocated
    to a given group $j$ is
    \be
    n_j = N_{\mathrm{new}} \left( \frac{w_j\sigma_j}{\sum_j w_j\sigma_j} \right) \,.
    \ee

    \item Within each segment $j$, evenly space across $\beta$ the new $n_j$ allocated samples.

    \item Collect $N_{\mathrm{new}}$ samples with this new $\beta$ schedule.

    \item Repeat steps 2--6 several times to refine the allocation of samples, decreasing
    the total variance of the estimator.

    \item Integrate the log posterior ratio of the samples across $\beta$ by summing with the weights $w_j$, to get the BF estimate.
\end{enumerate} 

In Section~\ref{sec: Toy Model}, we showcase this method with a toy model. In this simple case with only two parameters we can easily draw from the distribution using rejection sampling, but there are additional considerations to be made when sampling with MCMC.

\subsection{Relation to Steppingstone Sampling}

An alternative way to compute the Bayes factor is the steppingstone
(SS) method \citep{xie_et_al_2010,fan_rui_chen_kuo_lewis_2010,baele_lemey_vansteelandt_2013}.  This method involves
averaging a power of the posterior ratio (rather than averaging its
logarithm as in TI) and multiplying these averages over a series of steps.
Instead of~(\ref{eq:ms_ti}), the model-switch form of SS uses
\be
    \label{eq:ms_ss}
    \mathrm{BF}_{B,A}^{\mathrm{SS}} = \prod_{j=0}^{M-1} E_{\beta}
    \left[\left(\frac{L_B(d|\theta) \pi_B(\theta)}{L_A(d|\theta) \pi_A(\theta)}\right)^{\Delta\beta_j}\right] \,,
\ee
where the spacing between the $\beta$ values is given by
$\Delta\beta_j=\beta_{j+1}-\beta_j$.  Regardless of the number of
steps,~(\ref{eq:ms_ss}) is exact.  With $n$ samples at each $\beta_j$, we
have the estimator
\be
    \label{eq:ms_ss1}
    \widehat{\mathrm{BF}}_{B,A}^{\mathrm{SS}} = \prod_{j=0}^{M-1}
    \frac{1}{n}\sum_{i=1}^n\left(\frac{L_B(d|\theta_i) \pi_B(\theta_i)}{L_A(d|\theta_i) \pi_A(\theta_i)}\right)^{\Delta\beta_j} \,.
\ee
For every set of $\beta_j$, (\ref{eq:ms_ss1}) converges to the true
value as $n\to\infty$.  There is nothing corresponding to the
integration error of TI.  However, there is another source of
inaccuracy: one of the posterior ratios may be extremely large and
dominate the others.  This is not a problem in (\ref{eq:ms_ti_estimate})
because of the logarithm.  To reduce this inaccuracy, we should increase
the number of $\beta$ values and so decrease the $\Delta\beta_j$.
Then the large ratio is taken to a tiny power and so does not make
an unusual contribution.

When $\Delta\beta_j\ll1$, however, we have
\be
(\cdots)^{\Delta\beta_j} \approx 1+\Delta\beta_j\ln(\cdots) \,,
\ee
so
\be
\left\langle(\cdots)^{\Delta\beta_j}\right\rangle
\approx 1 + \Delta\beta_j \left\langle\ln(\cdots)\right\rangle
\approx \exp\left(\Delta\beta_j \left\langle\ln(\cdots)\right\rangle\right) \,,
\ee
and
\be
\mathrm{BF}_{B,A}^{\mathrm{SS}}
\approx \exp\sum_j \Delta\beta_j \, E_{\beta_j} \left[\ln\frac{L_B(d|\theta)
  \pi_B(\theta)}{L_A(d|\theta) \pi_A(\theta)}\right] \,,
\ee
which is just a discrete approximation to the integral of~(\ref{eq:ms_ti}).

Thus, to increase the accuracy of either TI or SS, we should use
closely spaced $\beta_j$, and then the two methods become the same.

\section{Toy Model}
\label{sec: Toy Model}
In this section, we test the procedure above on a toy model that is simple enough that the true BF can be computed directly (using a numerical integration) to any desired accuracy and that we can directly draw samples of the posterior for any $\beta$ (using rejection sampling).  The computations are fast enough that we can do many runs of the analysis to determine the spread of values that it returns.

Our toy model is inspired by analyses that compute BFs of different gravitational wave models performed by the North American Nanohertz Observatory for Gravitational Waves (NANOGrav), where we are members (see, e.g., \citet{agazie_2023})\footnote{For a gravitational wave background analysis using the steppingstone method, see \citet{zahraoui_et_al_2025}}. For the toy model we imagine two observations $x$ and $y$.  In Model A there is a single Gaussian background signal that might give rise to these observations and a single parameter $s_1$ giving the signal strength, the same for $x$ and $y$.  Model B includes that same signal but also an additional Gaussian signal with strength $s_2$ in $x$ but $10 s_2$ in $y$. We imagine that we have made 100 observations of the values $x$ and $y$ arising independently according to the model, and we would like to determine how much this dataset prefers one model over the other.  Within the context of model-switch, we define Model A to be the $\beta=0$ side of the path and Model B to be the $\beta=1$ side. The parameters are $\theta = (s_1, s_2)$, but the Model A likelihood does not depend on $s_2$.

\subsection{Model A (1 parameter)}

We draw $x$ and $y$ data independently from a
Gaussian with variance $s_1$. Then, each is subjected to additional Gaussian
noise of width $n$, so the total variance of each datum is $n^2+s_1$. The
likelihood of such a model is
\be\label{eq:LA}
L_A = \frac{1}{\sqrt{|2\pi\Sigma|}}
\exp\left(-\frac{1}{2}d^T\Sigma^{-1}d\right) \,,
\ee
with $d$ a 1D array of twice the length of the data points,
\be\label{eq:d}
d = [x_1, y_1, x_2, y_2, \ldots],
\ee
and $\Sigma$ a block-diagonal matrix,
\be\label{eq:Sigma}
\Sigma =
\begin{bmatrix}
\sigma & 0 & \cdots \\
0 & \sigma & \cdots \\
\vdots & \vdots & \ddots
\end{bmatrix} \,,
\ee
with
\be
\sigma =
\begin{bmatrix}
n^2+s_1 & 0 \\
0 & n^2+s_1
\end{bmatrix} \,.
\ee

By defining
\be
q = \sum_{i=1}^N \left(x_i^2+y_i^2\right) \,,
\ee
we can write the likelihood of model A with respect to N data as
\be
L_A =
\frac{1}{\left(2\pi(n^2+s_1)\right)^N}
\exp\left(
-\frac{1}{2}\frac{q}{n^2+s_1}
\right) \,.
\ee

To compute the marginal likelihood, we will use a flat prior $\pi(s_1) = 1/100$ for $0\le s_1 \le 100$ to simplify the model, and integrate the likelihood,
\be
Z_A = \frac{(2\pi)^{-N}}{10^2}
\int_0^\infty
\frac{1}{(n^2+s_1)^N}
\exp\left(
-\frac{1}{2}\frac{q}{n^2+s_1}
\right)\,ds_1 \,.
\ee

\subsection{Model B (2 parameters)}

We draw $x$ data values from a Gaussian with variance
$s_1+s_2$, while drawing $y$ independently from a Gaussian with
variance $s_1+10s_2$. Then each is subjected to additional Gaussian noise of
width $n$, so the total variances are $n^2+s_1+s_2$ for $x$ and
$n^2+s_1+10s_2$ for $y$. The likelihood is just as in~(\ref{eq:LA})
with the same data vector $d$ and form of the block-diagonal covariance matrix~(\ref{eq:Sigma}),
where now
\be
\sigma =
\begin{bmatrix}
n^2+s_1+s_2 & 0 \\
0 & n^2+s_1+10s_2
\end{bmatrix} \,.
\ee

By defining
\be
X = \sum_{i=1}^N x_i^2
\qquad \text{and} \qquad
Y = \sum_{i=1}^N y_i^2 \,,
\ee
we can write the likelihood of model B with respect to N data as 
\be
L_B =
\frac{\exp\left[
-\frac{1}{2}
\left(
\frac{X}{n^2+s_1+s_2}
+
\frac{Y}{n^2+s_1+10s_2}
\right)
\right]}{
\left[(2\pi)^2(n^2+s_1+s_2)(n^2+s_1+10s_2)\right]^{N/2}} \,.
\ee

To compute the marginal likelihood, we will use a flat prior with $\pi(s_1, s_2)$ constant for $0 \le s_1\le 100$ and $0 \le s_2\le 100$ and integrate the likelihood
\be
Z_B = \frac{(2\pi)^{-N}}{10^4}
\int_0^\infty \int_0^\infty
\frac{\exp\left[
-\frac{1}{2}
\left(
\frac{X}{n^2+s_1+s_2}
+
\frac{Y}{n^2+s_1+10s_2}
\right)
\right]}{
\left[(n^2+s_1+s_2)(n^2+s_1+10s_2)\right]^{N/2}}
\,ds_1\,ds_2 \,.
\ee

\begin{figure}
   \centering
    \includegraphics[width=1\linewidth]{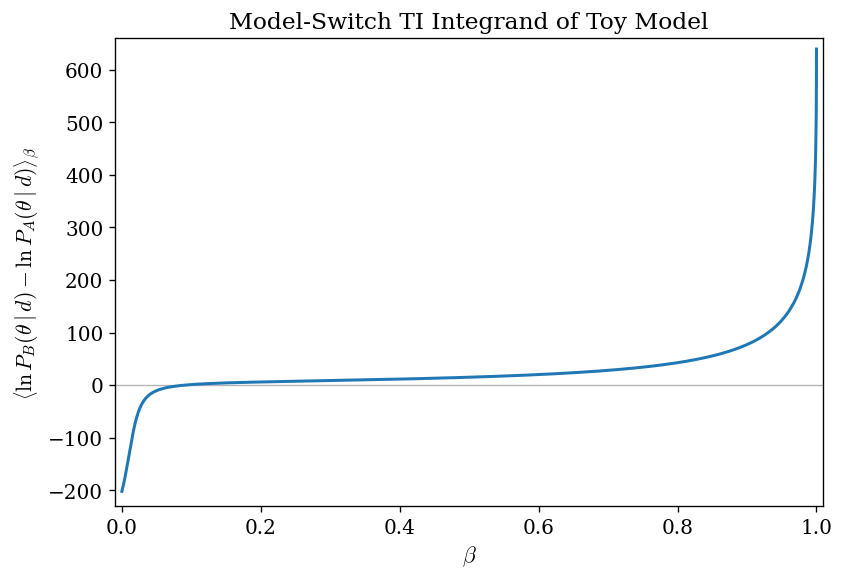}
    \caption{The integrand for the model-switch TI of the toy model that we are approximating. $\beta =0$ is Model A and $\beta =1$ is Model B. The integral of this function is $\ln \mathrm{BF}_{B,A}$. The integrand starts negative, making a negative contribution to $\ln \mathrm{BF}_{B,A}$, but rises sharply, before leveling out above 0 between approximately $\beta=0.1$ and $\beta=0.8$. Shortly before $\beta=1$, the function rises sharply again, making a large positive contribution to $\ln \mathrm{BF}_{B,A}$. It is easy to see from this plot that the area under the curve will be positive so the Bayes factor will favor Model B.}
    \label{fig:contrib_BF}
\end{figure}

We intentionally created a hard problem to solve, mimicking the steep curves near $\beta=0$ and $\beta=1$ produced by NANOGrav data and models. Figure~\ref{fig:contrib_BF} depicts the challenging integrand that we are trying to approximate. The data used in the following analysis are drawn from Model B. We used N=100 data points, and from those data got X=204.90012 and 
Y=1186.06538. We compute a true $\ln \mathrm{BF}_{B,A}$ of 28.2019.

\section{Adaptive Model-Switch TI with the Toy Model}
\label{sec: Adaptive Model-Switch TI with the Toy Model}

\subsection{Independent Draws from the Posterior}
\label{subsec: Independent Draws from the Posterior}
\begin{figure}
    \centering
    \includegraphics[width=1\linewidth]{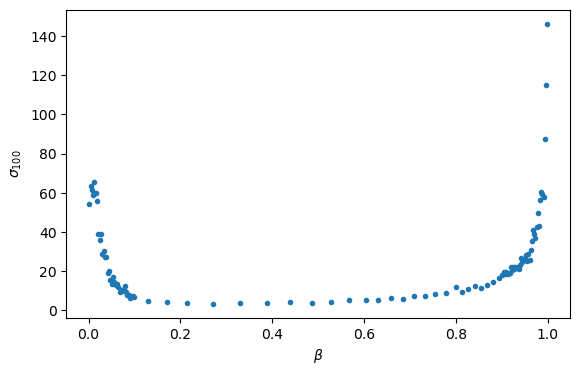}
    \caption{Standard deviation of the sampled log posterior differences as a function of $\beta$ for our model-switch toy model. This example is only one of the 100 iterations run for the $10^4$ $\beta$ schedule. The standard deviation is estimated by groups of 100 consecutive samples. Areas near $\beta =0$ and $\beta =1$ have the highest standard deviations, but rapidly fall off in the middle of the $\beta$ schedule, suggesting that more samples should be allocated towards the ends of the $\beta$ schedule. As we iterate, adding more $\beta$ will give a better estimate of the standard deviation curve.}
    \label{fig:stdev}
\end{figure}

We start with $10^3$ evenly spaced $\beta$, with samples drawn via rejection sampling according to the unnormalized model-switch distribution $q_\beta$, where $q_0$ and $q_1$ are set as in~\eqref{eq:ms_unnormalized}. A total of $10^3$ $\beta$ is sufficient to capture the overall shape of the integrand while remaining computationally inexpensive. It also allows us to subdivide the samples into 10 groups of 100 samples, which we use to construct the standard deviation curve that determines how new samples are allocated, as seen in Figure~\ref{fig:stdev}.
For each $\beta$ we draw two samples and average them. This is to emulate the forward and backward runs required when sampling with MCMC, described in Section~\ref{subsec: Draws Using MCMC}. Thus the total number of samples used is twice the number of $\beta$, for better comparison across sampling methods. Regions with larger standard deviation will receive a higher density of samples in subsequent runs. 

\begin{figure}[hbtp]
    \centering
    \includegraphics[width=1\linewidth]{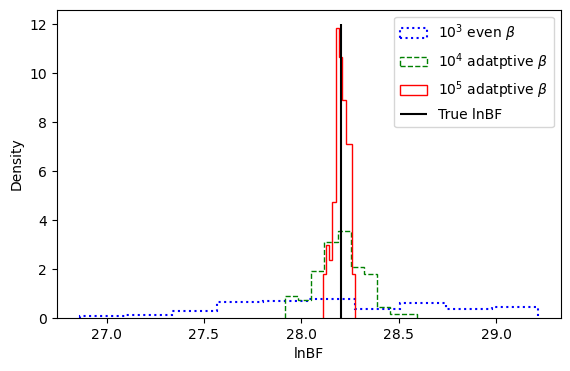}
    \caption{Distribution of the $\ln \mathrm{BF}$s produced by 100 iterations of adaptive TI with samples drawn from the posterior via rejection sampling. We show the distributions of the initial $10^3$ evenly spaced $\beta$ (dotted blue) and the two subsequent adaptive runs, with $10^4$ $\beta$ (dashed green) and $10^5$ $\beta$ (solid red). The true value of the $\ln \mathrm{BF}$ for the toy model is shown with a vertical solid black line at 28.2019. The mean and standard deviations of these distributions are below, along with the improvement in standard deviation produced by using more and adaptively spaced $\beta$. We also include the mean of errors we estimated from individual iterations.}
    \vspace{2ex}
    \begin{tabular}{c|c|c|c}
     \# of $\beta$ & \shortstack[2]{Recovered\\$\ln \mathrm{BF}$ values} & \shortstack[2]{Gain from\\previous line} & \shortstack[2]{Typical\\estimated error}\\
     \hline
        $10^3$ & 28.193 $\pm$ 0.542 &  & 1.1036\\
        $10^4$ & 28.199 $\pm$ 0.124 & 4.371 & 0.137\\
        $10^5$ & 28.202 $\pm$ 0.0361 & 3.435 & 0.0361
    \end{tabular}
    \label{fig:BF_Hist_RS}
\end{figure}

We ran the adaptive procedure twice, increasing the total number of samples by a factor of 10 at each stage, and subdividing into groups of 100 samples for standard deviation computation at each stage, ending with $10^5$ samples. Because these are random draws, we performed 100 independent iterations of the entire procedure to characterize the resulting distribution of $\ln \mathrm{BF}$ estimates. For every set of samples taken, we compute the $\ln \mathrm{BF}$ using~(\ref{eq:ms_ti_estimate}), producing a distribution of $\ln \mathrm{BF}$ values.  Repeating the entire procedure 100 times allows us to estimate the accuracy of our result. The resulting distributions are shown in Figure~\ref{fig:BF_Hist_RS}, illustrating convergence as the number of samples increases. 

\begin{figure}[hbtp]
    \centering
    \includegraphics[width=1\linewidth]{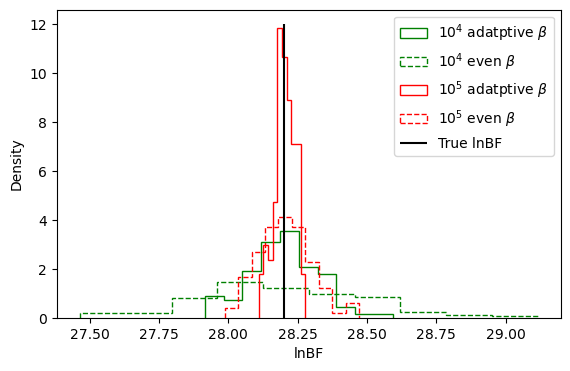}
    \caption{Distribution of the $\ln \mathrm{BF}$s produced by 100 iterations of adaptively spaced $\beta$ in comparison with 100 iterations of evenly spaced $\beta$, with samples drawn from the posterior via rejection sampling. The broader solid green distribution is $10^4$ adaptively spaced $\beta$ and the narrower solid red distribution is $10^5$ adaptively spaced $\beta$. The broader dashed green distribution is $10^4$ evenly spaced $\beta$ and the narrower dashed red distribution is $10^5$ evenly spaced $\beta$. The true value of the $\ln \mathrm{BF}$ for the toy model is shown with a vertical black line. The mean and standard deviations of these distributions are below, along with the improvement in standard deviation from using an adaptive $\beta$ schedule.}
    \vspace{2ex}
    \begin{tabular}{c|c|c|c}
     Spacing & \# of $\beta$ & \shortstack[2]{Recovered\\$\ln \mathrm{BF}$ values} & \shortstack[2]{Gain from\\previous line} \\
     \hline
        Even & $10^4$ & 28.21 $\pm$ 0.299 & \\
        Adaptive & $10^4$ & 28.199 $\pm$ 0.124 & 2.411\\
        Even & $10^5$ & 28.207 $\pm$ 0.097 & \\
        Adaptive & $10^5$ & 28.202 $\pm$ 0.0361 & 2.687
    \end{tabular}
    \label{fig:BF_Hist_RS_Adapt_Compare}
\end{figure}

We expect a $\sqrt{10}\approx 3.16$ factor of improvement in the standard deviation of the results just from increasing the number of samples by a factor of 10.  The use of adaptive sampling should lead to a even larger improvement.  Indeed going from $10^3$ evenly spaced samples to $10^4$ adaptively spaced samples gives a factor of 4.371 decrease in the spread, and going from $10^4$ to $10^5$ adaptive samples gives an improvement of 3.435.  In Figure~\ref{fig:BF_Hist_RS_Adapt_Compare} we compare the adaptive spacing of the $10^4$ and $10^5$ samples to their evenly spaced alternatives, finding factors of 2.411, and 2.687 improvement, respectively, in the standard deviation.
 
While the average $\ln \mathrm{BF}$s are fairly close to the true value of 28.2019, it is important to quantify the error associated with a single run, since application of this adaptive method will be for scenarios in which the true value is unknown and 100 iterations of sampling is not feasible. We can estimate the standard deviations reported beneath Figure~\ref{fig:BF_Hist_RS} with a single run fairly reliably, by computing the standard deviations of groups of 100 consecutive $\beta$ values, assuming that value to represent the sampling error of each $\beta$ in that group, weighting it by the size of the spacing, and adding in quadrature, to get a total standard deviation of the integral. For the $10^3$, $10^4$, and $10^5$ $\beta$ schedules, we compute the means across the 100 iterations of the individual errors to be 1.1036, 0.137, and 0.0361, respectively.

\subsection{Draws Using MCMC}
\label{subsec: Draws Using MCMC}
When the samples are generated by MCMC instead of independent draws, the relevant quantity is the effective sample size rather than the raw number of saved samples. Strong autocorrelation can make a seemingly dense schedule much less informative than it appears \citep{Geyer1992}. 

In this work we use an extended version of PTMCMCSampler \citep{justin_ellis_2017_1037579} that we modified to be able to perform model-switch thermodynamic integration and vary $\beta$ in accordance with a given $\beta$ schedule. This version of PTMCMCSampler will be made publicly available on GitHub upon publication of this paper. 

We start by computing chains with $\beta$ fixed at $0, 0.1, 0.2, ... 0.9, 1$, to get an idea of the appropriate burn-in length and the largest autocorrelation length of these runs. Once we acquire estimates for those values, we start two separate $\beta$ runs\footnote{For further parallelism, we could divide up the range of $\beta$ and run separate jobs to scan over each subrange. This decreases the real time to get the answer, but increases the total runtime, because each chain requires its own burn-in.}, one beginning at $\beta=0$ and increasing to $\beta=1$, and the other beginning at $\beta=1$ and decreasing to $\beta=0$. Averaging the forward and reverse traversals of the path helps avoid ``thermic lag'' \citep{lartillot_philippe_2006}, which occurs when the sampler is not entirely converged to the distribution appropriate to the $\beta$ for which it is sampling. After the appropriate burn-in period, we increment or decrement $\beta$ after every sample. 

If we vary $\beta$ too quickly, the resulting estimates may inherit bias from incomplete relaxation, so in this scenario, we started with a total of $10^5$ evenly spaced $\beta$ after a burn-in period. We estimate the autocorrelation length to be on the order of 100, this yields about $10^3$ effective (independent) samples in each direction. For each run we average the samples from the two opposing traversals at each $\beta$ before computing $\ln \mathrm{BF}$, so the true total number of samples is double that of the $\beta$ schedule. Then based on the standard deviation of the averaged samples, we propose a new adaptive $\beta$ schedule. Because we are now working with correlated samples, the standard deviation estimation is done by subdividing into groups of $10^4$ $\beta$ to take into account the autocorrelation length of approximately 100 and maintains the same number of groups per adaptive run as in Section~\ref{subsec: Independent Draws from the Posterior}.

\begin{figure}[hbtp]
    \centering
    \includegraphics[width=1\linewidth]{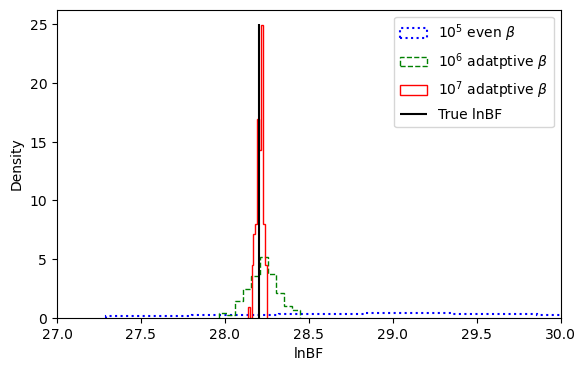}
    \caption{As in Figure~\ref{fig:BF_Hist_RS} with samples drawn via MCMC.  The table includes the number of $\beta$ values actually used in the MCMC and the effective number after dividing by the autocorrelation length.}
    \vspace{2ex}
    \begin{tabular}{c|c|c|c|c}
     \shortstack[c]{\# of\\raw $\beta$} & \shortstack[c]{\# of\\effective} $\beta$ & \shortstack[2]{Recovered\\$\ln \mathrm{BF}$ values} & \shortstack[2]{Gain from \\previous line} & \shortstack[2]{Typical\\estimated error}\\
     \hline
        $10^5$ & $10^3$ & 29.322 $\pm$ 1.077 &  &  1.164 \\
        $10^6$ & $10^4$ & 28.226 $\pm$ 0.0918 & 11.732 & 0.140\\
        $10^7$ & $10^5$ & 28.206 $\pm$ 0.0213 & 4.310 & 0.0360
    \end{tabular}
    \label{fig:BF_Hist_MCMC}
\end{figure}

In this study we use this procedure for $10^6$ adaptive $\beta$, and again with $10^7$ adaptive $\beta$, with effective sample sizes $10^4$ and $10^5$ respectively, as seen in Figure~\ref{fig:BF_Hist_MCMC}. The effective sample sizes allows us to reasonably compare to the independent draws in Section~\ref{subsec: Independent Draws from the Posterior}. We find similar results for the mean $\ln \mathrm{BF}$ and standard deviation over 100 iterations, showing that with the right measures the adaptive TI method works when using MCMC.  We estimate the error for each run as before, but now we combine the standard deviations of groups of $10^4$ $\beta$ values to again take account of the autocorrelation length. The mean of the individual errors of the $10^5$, $10^6$, and $10^7$ $\beta$ schedules are 1.164, 0.140, and 0.0360, respectively.

\section{Conclusion}
\label{sec: Conclusion}
In this paper we presented an adaptive model-switch thermodynamic integration method.  We used a $\beta$ schedule which allocates samples whose density in $\beta$ is proportional to the standard deviation of the samples in that area of $\beta$. We spread out the samples in $\beta$ to decrease discretization error and showed that this does not increase the overall variance as compared to gathering several samples at each $\beta$. We outlined the general procedure for implementing the adaptive $\beta$ schedule iteratively. 

We tested the method on a toy model with a known true BF value, using both independent posterior draws and MCMC sampling. In both cases, adaptive sampling increased accuracy of the results, which approached the known $\ln \mathrm{BF}$ as the effective sample size increased, thus supporting the feasibility and convergence of the adaptive MSTI procedure with either method. 

The results indicate that estimating the local variability of the MS integrand and concentrating $\beta$ values in high variance regions can use a fixed sampling budget more efficiently than even spacing.  Practical MCMC applications must also account for autocorrelation and effective sample size, burn-in, and possible lag while traversing the $\beta$ path. 

\section{Acknowledgments}
The authors acknowledge the Tufts University High Performance Compute Cluster, which was utilized for the research reported in this paper.

\section{Funding}
This work was supported in part by the National Science Foundation under Grant Nos. 2020265 and 2207267. 

\section{Disclosure Statement}
The authors report there are no competing interests to declare. 

\appendix
\section{Appendix}\label{sec: FDT}

In single-model TI, \cite{friel_hurn_wyse_2013} showed that the gradient of the expected log likelihood curve is equal to the variance of the log likelihood.  This is a general property arising from the fluctuation-dissipation theorem.  (See, e.g.~\cite{liu2024renormalization})  We show here that it applies to the model-switch case also.
Applying the second derivative to the fundamental TI equation \eqref{eqn: derivative TI},
\begin{align}
\label{eq: 2nd deriv var}
\frac{d}{d\beta} E_\beta \left[ \frac{d}{d\beta}\ln q_\beta(\theta)\right] &= \frac{d^2}{d\beta^2} \ln Z_\beta = \frac{d}{d\beta} \left( \frac{dZ_\beta/d\beta}{Z_\beta}\right) = \frac{d^2 Z_\beta/d\beta^2}{Z_\beta} - \left(\frac{dZ_\beta/d\beta}{Z_\beta}\right)^2 \nonumber\\
& = E_\beta \left[\frac{1}{q}\frac{d^2}{d\beta^2}q_\beta(\theta)\right]
- \left(E_\beta \left[\frac{d}{d\beta}\ln q_\beta(\theta)\right]\right)^2
\end{align}
Equation \eqref{eq: 2nd deriv var} holds for any $q_\beta$, but for $q_\beta$ linear in $\beta$, this reduces to the variance of the derivative of $\ln q_\beta$.
For the model-switch path, we have
\be
q_\beta(\theta) = q_0(\theta)^{1-\beta}q_1(\theta)^\beta
\ee
with $q_1(\theta) = L_B \pi_B$ and $q_0(\theta) = L_A \pi_A$.  Then \eqref{eq: 2nd deriv var} becomes
\be
\frac{d}{d\beta} E_\beta \left[\ln\left(\frac{q_1(\theta)}{q_0(\theta)}\right)\right]
= E_\beta \left[\left(\ln \frac{q_1(\theta)}{q_0(\theta)}\right)^2\right] - \left(E_\beta \left[\ln \frac{q_1(\theta)}{q_0(\theta)}\right]\right)^2  = \text{Var}_\beta \left[\ln\left(\frac{q_1(\theta)}{q_0(\theta)}\right)\right]\, .
\ee
This tells us that rapidly changing regions of the integrand correspond to higher variance, and suggests that more samples should be drawn from these regions for a more accurate integration.

\bibliography{ref.bib} 

@article{lartillot_philippe_2006, title={{Computing Bayes Factors Using Thermodynamic Integration}}, volume={55}, DOI={https://doi.org/10.1080/10635150500433722}, number={2}, journal={Systematic Biology}, author={Lartillot, Nicolas and Philippe, Hervé}, editor={Lewis, Paul}, year={2006}, month={Apr}, pages={195–207} }

@article{friel_pettitt_2008, title={{Marginal likelihood estimation via power posteriors}}, volume={70}, DOI={https://doi.org/10.1111/j.1467-9868.2007.00650.x}, number={3}, journal={Journal of the Royal Statistical Society: Series B (Statistical Methodology)}, author={Friel, N. and Pettitt, A. N.}, year={2008}, month={Jul}, pages={589–607} }

@article{kirkwood_1935, title={{Statistical Mechanics of Fluid Mixtures}}, volume={3}, DOI={https://doi.org/10.1063/1.1749657}, number={5}, journal={The Journal of Chemical Physics}, author={Kirkwood, John G.}, year={1935}, month={May}, pages={300–313} }

@article{neyman_1934, title={On the Two Different Aspects of the Representative Method: The Method of Stratified Sampling and the Method of Purposive Selection}, volume={97}, DOI={https://doi.org/10.2307/2342192}, number={4}, journal={Journal of the Royal Statistical Society}, author={Neyman, Jerzy}, year={1934}, pages={558} }

@article{lee_jahanmahin_pal_york_2025, title={{SAMTI: Sampling Adaptive Thermodynamic Integration for Alchemical Free Energy Calculations}}, volume={129}, DOI={https://doi.org/10.1021/acs.jpcb.5c05358}, number={51}, journal={The Journal of Physical Chemistry B}, publisher={American Chemical Society (ACS)}, author={Lee, Tai-Sung and Jahanmahin, Omid and Pal, Saikat and York, Darrin M.}, year={2025}, month={Dec}, pages={13063–13087} }

@article{lepage_2007, title={{A General Comparison of Relaxed Molecular Clock Models}}, volume={24}, DOI={https://doi.org/10.1093/molbev/msm193}, number={12}, journal={Molecular Biology and Evolution}, author={Lepage, T. and Bryant, D. and Philippe, H. and Lartillot, N.}, year={2007}, month={Jun}, pages={2669–2680} }

@article{Geyer1992,
  author  = {Geyer, Charles J.},
  title   = {{Practical Markov chain Monte Carlo (with discussion)}},
  journal = {Statistical Science},
  year    = {1992},
  volume  = {7},
  pages   = {473--511}
}

@article{aponte_2021, title={{An introduction to thermodynamic integration and application to dynamic causal models}}, volume={16}, DOI={https://doi.org/10.1007/s11571-021-09696-9}, number={1}, journal={Cognitive Neurodynamics}, author={Aponte, Eduardo A. and Yao, Yu and Raman, Sudhir and Frässle, Stefan and Heinzle, Jakob and Penny, Will D. and Stephan, Klaas E.}, year={2021}, month={Jul}, pages={1–15} }

@article{jeffreys_1935, title={{Some Tests of Significance, Treated by the Theory of Probability}}, volume={31}, DOI={https://doi.org/10.1017/s030500410001330x}, number={02}, journal={Mathematical Proceedings of the Cambridge Philosophical Society}, author={Jeffreys, Harold}, year={1935}, month={Apr}, pages={203} }

@book{jeffreys_1998, title={{The Theory of Probability}}, ISBN={9780191589676}, publisher={OUP Oxford}, author={Jeffreys, Harold}, year={1998}, month={Aug} }

@article{vitoratou_ntzoufras_2016, title={{Thermodynamic Bayesian model comparison}}, volume={27}, DOI={10.1007/s11222-016-9678-6}, number={5}, journal={Statistics and Computing}, publisher={Springer Science and Business Media LLC}, author={Vitoratou, Silia and Ntzoufras, Ioannis}, year={2016}, month={june}, pages={1165–1180} }

@article{agazie_2023, title={{The NANOGrav 15 yr Data Set: Evidence for a Gravitational-wave Background}}, volume={951}, DOI={10.3847/2041-8213/acdac6}, number={1}, journal={The Astrophysical Journal Letters}, author={Agazie, G Y and others}, year={2023}, month={june}, pages={L8-L8}, collaboration={NANOGrav}}

@misc{justin_ellis_2017_1037579,
  author       = {Justin Ellis and
                  Rutger van Haasteren},
  title        = {jellis18/PTMCMCSampler: Official Release},
  month        = oct,
  year         = 2017,
  doi          = {10.5281/zenodo.1037579},
  url          = {https://doi.org/10.5281/zenodo.1037579}
}

@article{xie_et_al_2010,
  author  = {Xie, Wangang and Lewis, Paul O. and Fan, Yu and Kuo, Lynn and Chen, Ming-Hui},
  title   = {{Improving Marginal Likelihood Estimation for Bayesian Phylogenetic Model Selection}},
  journal = {Systematic Biology},
  year    = {2011},
  month   = mar,
  volume  = {60},
  number  = {2},
  pages   = {150--160},
  doi     = {10.1093/sysbio/syq085}
}

@article{baele_lemey_vansteelandt_2013,
  author  = {Baele, Guy and Lemey, Philippe and Vansteelandt, Stijn},
  title   = {{Make the Most of Your Samples: Bayes Factor Estimators for High-Dimensional Models of Sequence Evolution}},
  journal = {BMC Bioinformatics},
  year    = {2013},
  month   = mar,
  volume  = {14},
  pages   = {85},
  doi     = {10.1186/1471-2105-14-85}
}

@article{zahraoui_et_al_2025,
  author  = {Zahraoui, El Mehdi and Maturana-Russel, Patricio and van Straten, Willem and Meyer, Renate and Gulyaev, Sergei},
  title   = {{Generalized Stepping-Stone Sampling: Efficient Marginal Likelihood Estimation in Gravitational Wave Analysis of Pulsar Timing Array Data}},
  journal = {Monthly Notices of the Royal Astronomical Society},
  year    = {2025},
  month   = jul,
  volume  = {540},
  number  = {4},
  pages   = {3818--3825},
  doi     = {10.1093/mnras/staf953}
}

@article{kadane_lazar_2004,
  author  = {Kadane, Joseph B. and Lazar, Nicole A.},
  title   = {{Methods and Criteria for Model Selection}},
  journal = {Journal of the American Statistical Association},
  year    = {2004},
  month   = mar,
  volume  = {99},
  number  = {465},
  pages   = {279--290},
  doi     = {10.1198/016214504000000269}
}

@article{fan_rui_chen_kuo_lewis_2010, title={{Choosing among Partition Models in Bayesian Phylogenetics}}, volume={28}, url={https://doi.org/10.1093/molbev/msq224}, number={1}, journal={Molecular Biology and Evolution}, publisher={Oxford Academic}, author={Fan, Yu and Rui, Wu and Chen, Ming-Hui and Kuo, Lynn and Lewis, Paul O}, year={2010}, month=aug, pages={523–532} }

@article{friel_hurn_wyse_2013, title={Improving power posterior estimation of statistical evidence}, volume={24}, DOI={10.1007/s11222-013-9397-1}, number={5}, journal={Statistics and Computing}, author={Friel, Nial and Hurn, Merrilee and Wyse, Jason}, year={2013}, month=may, pages={709–723} }

@article{GelmanMeng,
  title={Simulating Normalizing Constants: From Importance Sampling to Bridge Sampling to Path Sampling},
  author={Andrew Gelman and Xiao-Li Meng},
  journal={Statistical Science},
  year={1998},
  volume={13},
  number={2},
  pages={163--185},
  url={https://projecteuclid.org/journals/statistical-science/volume-13/issue-2/Simulating-normalizing-constants--from-importance-sampling-to--bridge/10.1214/ss/1028905934.full}
}

@article{CalderheadGirolami,
title = {Estimating Bayes factors via thermodynamic integration and population MCMC},
journal = {Computational Statistics \& Data Analysis},
volume = {53},
number = {12},
pages = {4028-4045},
year = {2009},
issn = {0167-9473},
doi = {https://doi.org/10.1016/j.csda.2009.07.025},
url = {https://www.sciencedirect.com/science/article/pii/S0167947309002722},
author = {Ben Calderhead and Mark Girolami},
}

@article{oates_papamarkou_girolami_2016, title={The Controlled Thermodynamic Integral for Bayesian Model Evidence Evaluation}, volume={111}, DOI={10.1080/01621459.2015.1021006}, number={514}, journal={Journal of the American Statistical Association}, author={Oates, Chris J and Papamarkou, Theodore and Girolami, Mark}, year={2016}, month=apr, pages={634–645} }

@misc{liu2024renormalization,
  author       = {Liu, Zhihang},
  title        = {Renormalization Group},
  year         = {2025},
  url          = {https://github.com/Liu-Zhihang/Renormalization-Group},
  note         = {Self-study notes on critical phenomena, phase transitions, scaling laws and universality classes}
}

@article{friel_wyse_review,
author = {Friel, Nial and Wyse, Jason},
title = {Estimating the evidence – a review},
journal = {Statistica Neerlandica},
volume = {66},
number = {3},
pages = {288-308},
doi = {https://doi.org/10.1111/j.1467-9574.2011.00515.x},
url = {https://onlinelibrary.wiley.com/doi/abs/10.1111/j.1467-9574.2011.00515.x},
eprint = {https://onlinelibrary.wiley.com/doi/pdf/10.1111/j.1467-9574.2011.00515.x},
year = {2012}
}

\end{document}